\documentclass[12pt,thmsa]{article}
\usepackage{amssymb}

\input tcilatex
\QQQ{Language}{American English}

\begin{document}

\begin{center}
\emph{The Principles of Self Creation Cosmology and its Comparison with
General Relativity}

Garth A Barber
\end{center}
\begin{center}
\textit{Kingswood Vicarage, Woodland Way, Tadworth, SURREY KT20 6NW, England.%
}

Tel: +44 01737 832164\qquad e-mail: garth.barber@virgin.net
\end{center}

\begin{center}
\textit{Abstract}
\end{center}

There are, at present, several gravitational and cosmological anomalies; 
the dark energy problem, the lambda problem, accelerating cosmological
expansion, the anomalous Pioneer spacecraft acceleration, a spin-up of the
Earth and an apparent variation of G observed from analysis of the evolution
of planetary longitudes. These conundrums may be resolved in the theory of 
Self Creation Cosmology, in which the Principle of Mutual Interaction subsumes 
both Mach's Principle and the Local Conservation of Energy. The theory is 
conformally equivalent to General Relativity in vacuo with the consequence that
predictions of the theory are identical with General Relativity in the
standard solar system experiments. Other observable local and cosmological
consequences offer an explanation for the anomalies above. The SCC universe
expands linearly in its Einstein Frame and it is static in its Jordan Frame;
hence, as there are no density, smoothness or horizon problems, there is no
requirement for Inflation. The theory determines the total density parameter
to be one third, and the cold dark matter density parameter to be two
ninths, yet in the Jordan frame the universe is similar to Einstein's
original static cylindrical model and spatially flat. Therefore there is no
need for a 'Dark Energy' hypothesis. As the field equations determine the
false vacuum energy density to be a specific, and feasibly small, value
there is no 'Lambda Problem'. Finally certain observations in SCC would
detect cosmic acceleration.

\begin{center}
\ \ 
\end{center}
\newpage
\section{Introduction}

\subsection{A Possible Problem}

This paper suggests that a series of disparate observations may now be
raising questions about General Relativity (GR). The new theory of Self
Creation Cosmology (SCC), (Barber, 2002), is presented as a viable
alternative, against which GR can be theoretically and experimentally
compared.

The cosmological problems, as widely reported in the literature, are as
follows: Firstly, although the familiar density, smoothness and horizon
problems of GR are at present resolved by Inflation, observations of galaxy
clustering and gravitational lensing seem to indicate the density parameter
to be only about a third of that required (Chae et al. 2002). Consequently
the standard model demands the existence of unknown 'Dark Energy' to make up
the missing mass (Chae et al. 2002). If some, or all, of this mass is
false vacuum energy then, secondly, there is a 'Lambda Problem' in which the
actual density of such energy is about 121 magnitudes smaller than theory
predicts (Efstathiou, 1995). Finally, 'cosmological acceleration'
(Perlmutter et al., 1999) has presented GR with a formidable problem, which
is being resolved by such suggestions as that of dark energy or a dynamic 
cosmological constant $\Lambda \left( t\right) $. (e.g. Vishwakarma, 2002)

Closer to home the Pioneer spacecraft appear to have an anomalous sunward
acceleration (Anderson et al. 2002a). There may be several explanations of
this acceleration and it may have several components, however, as it has
been observed a number of times, the excess over the General Relativity
acceleration 
\[
a_P=\left( 8.74\text{ }\pm \text{ }1.3\right) \text{ x }10^{-8}\text{ cm/sec}%
^{\text{2}} 
\]
is equal to $cH$ if $H=87$ $km.sec^{\text{-1}}/Mpc$. Therefore it might be
cosmological in nature.

A second anomaly as reviewed by Leslie Morrison and Richard Stephenson
[(Morrison and Stephenson, 1998), (Stephenson, 2003)] arises from the analysis of the length of
the day from ancient eclipse records. It is that in addition to the tidal
contribution there is a long-term component acting to decrease the length of
the day which equals 
\[
\vartriangle \text{T/day/cy}=-6\ \text{x }10^{-4}\text{ sec/day/cy.} 
\]
This component, which is consistent with recent measurements made by
artificial satellites, is thought to result from the decrease of the Earths
oblateness following the last ice age. Although this explanation certainly
merits careful consideration, and it is difficult to separate the various
components of the Earth's rotation, it is remarkable that this value $%
\vartriangle $T/day/cy is equal to $H$ if $H=67$ $km.sec^{\text{-1}}/Mpc$.
The question is why should this spinning up of the Earth's rotation have a
natural time scale of the order of the age of the universe rather than the
natural relaxation time of the Earths crust or the periodicity of the ice
ages? This anomaly also may therefore be cosmological rather than
geophysical in nature.

A third anomaly, which arises from the analysis of the residues of planetary
longitudes, reveals that the Gravitational constant appears to be varying at
a rate also of the order of Hubble's constant. An analysis [Krasinsky et
al., (1985)] rendered a problematic value for a variation in $G$ 
\[
\frac{\stackrel{.}{G}}G\approx +\left( 4\pm 0.8\right) .10^{-11}yr^{-1} 
\]
with a caveat that the sign might be reversed. This value of $\frac{%
\stackrel{.}{G}}G$ is equal to $H$ if $H=38$ $km.sec^{\text{-1}}/Mpc$, and
therefore it too may be cosmological in nature.

If these are indeed three observations of Hubble's constant, then their
values have a spread typical of other determinations of $H$ with an average
of $H=64$ $km.sec^{-1}/Mpc$ in good agreement with more orthodox methods.
Although there may well be other explanations for these anomalies it is
remarkable that they all approximate Hubble's constant.

The question that arises, if these three observations are a signal for $H$,
together with the cosmological problems mentioned above, is,
''Notwithstanding the empirical success of GR, is there a problem with the theory?''

\subsection{A Possible Solution}

It was shown in the earlier paper (Barber, 2002) that SCC predicts identical
outcomes as GR in the classical tests. Therefore the high precision tests,
which have vindicated GR over many decades, do not falsify SCC either (Barber, 2003a). 
However two other falsifiable experiments have been proposed in principle (Barber, 2002), which 
are able to distinguish between the two theories. They ask the questions, ''Do photons fall at 
the same rate as particles?'' and ''Is there a cut off for the Casimir force which approaches 
zero as space-time curvature approaches flatness?'' Hence the theory is falsifiable. Furthermore 
there is another definitive test that is about to be performed on the Gravity Probe B satellite. 
Whereas SCC predicts a 'frame-dragging' result equal to GR it predicts a geodetic precession of 
only 5/6,or 0.83 the GR value (Barber, 2003a). Apart from these tests of GR this paper will show 
that SCC offers solutions to the cosmological and local conundrums described above. 

It will be seen that not only does the theory not suffer from a density, smoothness 
or horizon problem and therefore it does not require Inflation, but also the theory 
determines the universe's density to actually be a third of the critical density and 
therefore it does not require dark energy either. SCC determines the density of the 
false vacuum from its field equations to be a specific and feasibly small value,
thus it also appears to resolve the 'Lambda Problem'. Finally observations of distant standard 
candles such as SN Ia would detect cosmic acceleration.

The theory actually predicts that the Pioneer spacecraft would appear to experience a sunward 
acceleration equal to $cH$ because of a drift between atomic clock and ephemeris time. It also 
predicts rotating bodies should spin up at a rate equal to $H$. Finally it predicts 
$\frac{\stackrel{.}{G}}G$ to be $H$ but $\frac{\stackrel{.%
}{\left( GM\right) }}{GM}$ to be zero. Although a signal deduced from the evolution of orbital 
longitudes would detect the latter of these two, such a signal would also suffer the above clock 
drift. This would be interpreted as $\frac{\stackrel{.}{G}}G$ = $H$ 

It is here suggested that Self Creation Cosmology is a viable alternative to General Relativity.

\subsection{ The Principles of SCC}

Einstein gave some consideration to two concepts, the local conservation of energy and Mach's 
Principle, which are not fully included in GR. At various times since the publication of Einstein's 
GR papers these concepts have been considered independently, in SCC they are considered together.

The first non-GR concept, the Local Conservation of Energy can be
appreciated by considering the conservation of four-momentum, $P^\nu $, of a
projectile in free fall, which is a fundamental property of any metric
theory such as GR as it necessarily follows from the equivalence principle.
As a consequence the energy or 'relativistic mass' of a particle, ($P^0$),
is not conserved, except when measured in a co-moving frame of reference, or
in the Special Relativity (SR) limit. In any metric theory a particle's rest
mass is necessarily invariant as it is mathematically identical to the norm
of the four-momentum vector. This requirement here defines the Einstein
frame (EF). (Note: In the Brans and Dicke theory (BD) the EF is that
conformal frame in which $G$ is constant.) The local non-conservation of
energy is a consequence of the fact that energy is not a manifestly
covariant concept, that is its value is relative to the inertial frame of
reference in which it is measured. As the equivalence principle does not
allow a preferred frame, there is no definitive value for energy in any
metric theory.

The second non-(fully)GR concept is Mach's Principle. This suggests that
inertial frames of reference should be coupled to the distribution of mass
and energy in the universe at large, hence one would actually expect there
to be a preferred frame, that is a frame in which the universe as a whole
might be said to be at rest, in which $P^0$ is conserved, in apparent
contradiction to the spirit of the equivalence principle. In fact such a
frame of reference does appear to exist, it is that in which the Cosmic
Background Radiation (CBR) is globally isotropic.

These two problems are linked and resolved together in the new theory, SCC, 
by the proposal that energy is locally conserved when measured in a particular,
preferred, frame of reference as selected by Mach's principle, that is the
Center of Mass (CoM) of the system. It thus defines what is called the
Jordan (energy) Frame [JF(E)] in which rest mass is required to include
gravitational potential energy, as defined in that CoM frame of reference.

This local conservation of energy requires the energy expended in lifting an
object against a gravitational field to be translated into an increase in
rest mass. If $\Phi _N\left( x^\mu \right) $ is the dimensionless Newtonian
gravitational potential defined by a measurement of acceleration in a local
experiment in a frame of reference co-moving with the Centre of Mass frame
(CoM), 
\begin{equation}
\frac{d^2r}{dt^2}=-\nabla \Phi _N\left( r\right) \text{ }  \label{eq105}
\end{equation}
and normalized so that $\Phi _N\left( \infty \right) $ $=0$ , then the local
conservation of energy requires 
\begin{equation}
\frac 1{m_p\left( x^\mu \right) }\mathbf{\nabla }m_p\left( x^\mu \right) =%
\mathbf{\nabla }\Phi _N\left( x^\mu \right) \text{ ,}  \label{eq66}
\end{equation}
where $m_p(x^\mu )$ is measured locally at $x^\mu $. This has the solution 
\begin{equation}
m_p(x^\mu )=m_0\exp [\Phi _N\left( x^\mu \right) ]\text{ ,}  \label{eq24}
\end{equation}
\[
\text{where}\qquad m_pr\rightarrow m_0\qquad \text{as\qquad }r\rightarrow
\infty \text{ .} 
\]
The gravitational field equations of the new theory are modified to
explicitly include Mach's principle, following BD, (Brans \& Dicke, 1961),
by including the energy-momentum tensor of a scalar field energy $T_{\phi
\,\mu \nu }$ 
\begin{equation}
R_{\mu \nu }-\frac 12g_{\mu \nu }R=\frac{8\pi }\phi \left[ T_{M\mu \nu
}+T_{\phi \,\mu \nu }\right] \text{ .}  \label{eq4}
\end{equation}
where $T_{M\mu \nu }$ is the energy momentum tensors describing the matter
field. The scalar field $\phi \approx \frac 1{G_N}$ is coupled to the large
scale distribution of matter in motion, described by a field equation of the
simplest general covariant form 
\begin{equation}
\Box \phi =4\pi T_M^{\;}\text{ ,}  \label{eq2}
\end{equation}
$T_{M\;}^{\;}$ is the trace, ($T_{M\;\sigma }^{\;\;\sigma }$), of the energy
momentum tensor describing all non-gravitational and non-scalar field energy
and where the Brans Dicke parameter $\lambda $ has been determined to be
unity. (Barber, 2002)

In SCC mass is created out of the gravitational and scalar fields according
to the Principle of Mutual Interaction (PMI), in which the scalar field is a
source for the matter-energy field if and only if the matter-energy field is
a source for the scalar field.

\begin{equation}
\nabla _\mu T_{M\;\nu }^{.\;\mu }=f_\nu \left( \phi \right) \Box \phi =4\pi
f_\nu \left( \phi \right) T_{M\;}^{\;}\text{ ,}  \label{eq7}
\end{equation}
As a consequence photons still do traverse null-geodesics\textit{, }at least 
\textit{in vacuo,} \textit{\ } 
\begin{equation}
\nabla _\mu T_{em\quad \nu }^{\quad \mu }=4\pi f_\nu \left( \phi \right)
T_{em}^{\;}=4\pi f_\nu \left( \phi \right) \left( 3p_{em}-\rho _{em}\right)
=0  \label{eq8}
\end{equation}
where $p_{em}$ and $\rho _{em}$ are the pressure and density of an
electromagnetic radiation field with an energy momentum tensor $T_{em\,\mu
\nu }$ and where $p_{em}=\frac 13\rho _{em}$ and where it will be shown that
\begin{equation}
f_\nu \left( \phi \right) =\frac 1{8\pi \phi }\nabla _\nu \phi \text{ .}
\label{eq7a}
\end{equation}

\section{The SCC Conformal Transformation}

These SCC principles have the consequence (Barber, 2002) that in the Jordan
Frame, in which energy is locally conserved in the Centre of Mass frame of
reference, a photon has constant frequency and its energy is conserved even
when crossing a gravitational field. Gravitational red shift is
interpreted as a gain of potential energy, and hence mass, of the measuring
apparatus, rather than the loss of (potential) energy by the photon.

There are two questions to ask in order that a Weyl metric may be set up
spanning extended space-time; ''What is the invariant standard by which
objects are to be measured?'' and ''How is that standard to be transmitted
from event to event in order that the comparison can be made?'' In GR and
the SCC EF the principle of energy-momentum conservation, i.e. invariant
rest mass, determines that standard of measurement to be fixed rulers and
regular clocks. In the SCC JF(E), on the other hand, the principle of the
local conservation of energy determines that standard of measurement to be a
''standard photon'', with its frequency (inverse) determining the standard
of time and space measurement, and its energy determining the standard of
mass, all defined in the CoM, Machian, frame of reference.

In this theory the EF is the natural frame in which to interpret experiments
and observations of matter and the JF(E) is the natural frame in which to
interpret astronomical and cosmological observations and gravitational
orbits. The conformal transformation of a metric $g_{\mu \nu }$ into a
physically equivalent alternative $\widetilde{g}_{\mu \nu }$ is described by 
\begin{equation}
g_{\mu \nu }\rightarrow \widetilde{g}_{\mu \nu }=\Omega ^2g_{\mu \nu }\text{
.}  \label{eq9}
\end{equation}

The JF of SCC requires mass creation, ( $\nabla _\mu T_{M\;\nu }^{\;\mu
}\neq 0$ ), therefore the scalar field is non-minimally connected to matter.
The JF Lagrangian density is, 
\begin{equation}
L^{SCC}[g,\phi ]=\frac{\sqrt{-g}}{16\pi }\left( \phi R-\frac \omega \phi
g^{\mu \nu }\nabla _\mu \phi \nabla _\nu \phi \right)
+L_{matter}^{SCC}[g,\phi ]\text{ ,}  \label{eq24a}
\end{equation}
and its conformal dual, [Dicke (1962)], by a general transformation $%
\widetilde{g}_{\mu \nu }=\Omega ^2g_{\mu \nu }$ , is 
\begin{eqnarray}
L^{SCC}[\widetilde{g},\widetilde{\phi }] &=&\frac{\sqrt{-\widetilde{g}}}{%
16\pi }\left[ \widetilde{\phi }\widetilde{R}+6\widetilde{\phi }\widetilde{%
\Box }\ln \Omega \right] +\widetilde{L}_{matter}^{SCC}[\widetilde{g},%
\widetilde{\phi }]  \label{eq24b} \\
&&\ -\frac{\sqrt{-\widetilde{g}}}{16\pi }\left[ 2\left( 2\omega +3\right) 
\frac{\widetilde{g}^{\mu \nu }\widetilde{\nabla }_\mu \Omega \widetilde{%
\nabla }_\nu \Omega }{\Omega ^2}+4\omega \frac{\widetilde{g}^{\mu \nu }%
\widetilde{\nabla }_\mu \Omega \widetilde{\nabla }_\nu \widetilde{\phi }}%
\Omega +\omega \frac{\widetilde{g}^{\mu \nu }\widetilde{\nabla }_\mu 
\widetilde{\phi }\widetilde{\nabla }_\nu \widetilde{\phi }}{\widetilde{\phi }%
}\right] \text{ .}  \nonumber
\end{eqnarray}
Now mass is conformally transformed according to 
\begin{equation}
m\left( x^\mu \right) =\Omega \widetilde{m}_0  \label{eq33}
\end{equation}
[see Dicke, (1962)], where $m\left( x^\mu \right) $ is the mass of a
fundamental particle in the JF and $\widetilde{m}_0$ its invariant mass in
the EF. Therefore the local conservation of energy in the SCC JF, Equations 
\ref{eq24} and \ref{eq33}, require 
\begin{equation}
\Omega =\exp \left[ \Phi _N\left( x^\mu \right) \right] \text{ .}
\label{eq34}
\end{equation}

The question is, ''How does $\phi $ transform?'' In the BD EF and the GR JF
where gravitation and mass are inextricably combined, the conformal
transformation of the scalar field depends on the dimensionless and
therefore invariant, 
\begin{equation}
Gm^2=\widetilde{G}\widetilde{m}^2  \label{eq34e}
\end{equation}
\[
\text{i.e. }\widetilde{\phi }_{BD}=\phi _{BD}\Omega ^{-2}\text{ .} 
\]
Defining the conformal transformation $\Omega $ by 
\begin{equation}
\Omega =\left( G\phi \right) ^\alpha  \label{eq34d}
\end{equation}
then 
\begin{equation}
\widetilde{\phi }_{BD}=G^{-2\alpha }\phi _{BD}^{\left( 1-2\alpha \right) }%
\text{ ,}  \label{eq34a}
\end{equation}
which in the BD case, where $\widetilde{G}$ is constant, requires $\alpha
=\frac 12$.

In SCC, however, it is postulated that potential energy should also be
convoluted with gravitation and mass. This is achieved by including the
conformal parameter, $\Omega $, which is now an expression of potential
energy, with the gravitational 'constant' and mass. The dimensionless
conformal invariant now becomes 
\begin{equation}
Gm^2\Omega ^\beta =\widetilde{G}\widetilde{m}^2\widetilde{\Omega }^\beta 
\text{ .}  \label{eq34f}
\end{equation}
Now $\ \widetilde{\Omega }=1$ by definition therefore $G\widetilde{m}^2$ is
invariant in that frame, and as $\widetilde{m}$ is constant, hence $%
\widetilde{G}$ and consequentially $\widetilde{\phi }$ are constant. In this
case Equation \ref{eq34d} yields 
\begin{equation}
\widetilde{\phi }_{SCC}=G^{-\alpha \left( 2+\beta \right) }\phi
_{SCC}^{\left[ 1-\alpha \left( 2+\beta \right) \right] }\text{ ,}
\label{eq34b}
\end{equation}
which sets the following condition for the EF 
\begin{equation}
\beta =\frac 1\alpha -2\text{ .}  \label{eq34c}
\end{equation}
If $\omega =-\frac 32$ and $\widetilde{\Box }\ln \Omega =0$, then as $%
\widetilde{\phi }$ is constant, Equation \ref{eq24b} reduces to 
\begin{equation}
L^{SCC}[\widetilde{g}]=\frac{\sqrt{-\widetilde{g}}}{16\pi G_N}\widetilde{R}+%
\widetilde{L}_{matter}^{SCC}[\widetilde{g}]\text{ ,}  \label{eq24c}
\end{equation}
where matter is now minimally connected. Thus with these three conditions
the conformal transformation of the Lagrangian density, Equation \ref{eq24a}%
, reduces to canonical GR. The value $\omega =-\frac 32$ can either be set empirically 
(Barber, 2002) or determined from the first principles of the theory (Barber, 2003b).
This unique frame is designated the Jordan energy
Frame, [JF(E)] because in it energy is locally conserved. The last
condition, $\widetilde{\Box }\ln \Omega =0$, is the vacuum condition, $%
\widetilde{\Box }\widetilde{\Phi }_N\left( \widetilde{x}^\mu \right) =0$, as
this reduces to $\widetilde{\nabla }^2\widetilde{\Phi }_N\left( \widetilde{x}%
^\mu \right) =0$ in a harmonic coordinate system. The metrics thus relate 
\underline{in vacuo} according to Equation \ref{eq9} 
\begin{equation}
g_{\mu \nu }\rightarrow \widetilde{g}_{\mu \nu }=\exp \left[ 2\Phi _N\left(
x^\mu \right) \right] g_{\mu \nu }\text{ ,}  \label{eq35}
\end{equation}
where $\widetilde{g}_{\mu \nu }$ is the GR metric. As matter is minimally
connected in the EF it is necessary first to carry out the variational
principle in that frame and then conformally transform the result into the
JF(E).

The energy-momentum tensor of matter is thereby defined in the EF by 
\begin{equation}
\widetilde{T}_{M\mu \nu }^{SCC}=\frac 2{\sqrt{-\widetilde{g}}}\frac \partial
{\partial \widetilde{g}^{\mu \nu }}\left( \sqrt{-\widetilde{g}}\widetilde{L}%
_{matter}^{SCC}\right) \text{ .}  \label{eq25}
\end{equation}
The corresponding energy-momentum tensor of matter in the JF(E) is defined
by the conformal dual of this definition in the EF, 
\begin{equation}
\widetilde{T}_{M\;\mu \nu }^{SCC}\left( \widetilde{g}^{\mu \nu }\right)
\rightarrow T_{M\;\mu \nu }^{SCC}\left( g^{\mu \nu }\right) \text{ , where }%
g^{\mu \nu }=\exp \left[ 2\Phi _N\left( x^\mu \right) \right] \widetilde{g}%
^{\mu \nu }\text{.}  \label{eq27}
\end{equation}
The Lagrangian density in the EF is given by 
\begin{equation}
L^{SCC}[\widetilde{g},\widetilde{\phi }]=\frac{\sqrt{-\widetilde{g}}}{16\pi
G_N}\widetilde{R}+\widetilde{L}_{matter}^{SCC}[\widetilde{g}]+\frac{3\sqrt{-%
\widetilde{g}}}{8\pi G_N}\widetilde{\square }\widetilde{\Phi }_N\left( 
\widetilde{x}^\mu \right) \text{ .}  \label{eq25a}
\end{equation}
Its conformal dual in the JF(E) is that of Equation \ref{eq24a} with $\omega
=-\frac 32$, 
\begin{equation}
L^{SCC}[g,\phi ]=\frac{\sqrt{-g}}{16\pi }\left( \phi R+\frac 3{2\phi }g^{\mu
\nu }\nabla _\mu \phi \nabla _\nu \phi \right) +L_{matter}^{SCC}[g,\phi ]%
\text{ ,}  \label{eq26}
\end{equation}
and the corresponding field equation to this Lagrangian density is 
\begin{eqnarray}
R_{\mu \nu }-\frac 12g_{\mu \nu }R &=&\frac{8\pi }\phi T_{M\mu \nu }-\frac
3{2\phi ^2}\left( \nabla _\mu \phi \nabla _\nu \phi -\frac 12g_{\mu \nu
}g^{\alpha \beta }\nabla _\alpha \phi \nabla _\beta \phi \right) 
\label{eq29} \\
&&\ \ \ \ +\frac 1\phi \left( \nabla _\mu \nabla _\nu \phi -g_{\mu \nu }\Box
\phi \right) \text{ .}  \nonumber
\end{eqnarray}
The SCC EF Lagrangian is de-coupled from the scalar field. The novel feature
of SCC, distinguishing it from simple JF GR, is that Mach's Principle is
fully incorporated in the JF(E) by applying Equation \ref{eq2}. The
relationship between the field equations, \ref{eq2} and \ref{eq29} is
obtained by covariantly differentiating Equation \ref{eq29}. After
multiplying through by $\phi $ ($\neq 0$), taking $\nabla _\mu $, using the
Bianchi identities, Equation \ref{eq2} and the identity 
\[
\nabla _\sigma \phi R_{\;\nu }^\sigma =\nabla _\nu \left( \Box \phi \right) 
-\Box \left( \nabla _\nu \phi \right)\text{ ,}
\]
one obtains the PMI expression: 
\begin{equation}
\nabla _\mu T_{M\nu }^{\;\mu }=\frac 1{8\pi }\frac 1\phi \nabla _\nu \phi
\Box \phi \text{ .}  \label{eq31}
\end{equation}
On using Equation \ref{eq2} this becomes 
\begin{equation}
\nabla _\mu T_{M\nu }^{\;\mu }=\frac 12\frac 1\phi \nabla _\nu \phi T_M\text{
.}  \label{eq32}
\end{equation}
In this theory where the conformal invariant is $Gm^2\Omega ^\beta $ the
relationship between the scalar field in the JF(E) and EF is $\phi
_{SCC}=\Omega ^{\left( 2+\beta \right) }\widetilde{\phi }_{SCC}$. The
parameter $\beta $ is determined by the principle of the conservation of
energy. Furthermore when $\nabla _\mu \phi =0$ Equation \ref{eq32} reduces
to 
\begin{equation}
\nabla _\mu T_{M\nu }^{\;\mu }=0  \label{eq32a}
\end{equation}
and it was shown that in this immediate locality SCC reduces to SR. In that
locality, where $g_{\mu \nu }\rightarrow \eta _{\mu \nu }$ and $\phi =$
constant, the theory admits a ground state solution.

The field equation can be cast in a form that does not contain the
second derivatives of $\phi $, which are necessarily convoluted with the
gravitational terms of the affine connection. When cast into its 'corrected'
form the left hand side of the gravitational field equation, the Einstein
tensor $G_{\mu \nu }$ , becomes the 'affine' Einstein tensor $^\gamma G_{\mu
\nu }$ and in this case the whole JF(E) equation becomes 
\begin{equation}
^\gamma G_{\mu \nu }=\frac{8\pi }\phi T_{M\mu \nu }+\frac{\left( \omega
+\frac 32\right) }{\phi ^2}\left( \nabla _\mu \phi \nabla _\nu \phi -\frac
12g_{\mu \nu }g^{\alpha \beta }\nabla _\alpha \phi \nabla _\beta \phi \right)
\label{eq37}
\end{equation}
so in the SCC case, where $\omega =-\frac 32$, this reduces down to 
\begin{equation}
^\gamma G_{\mu \nu }=\frac{8\pi }\phi T_{M\mu \nu }\text{ .}  \label{eq38}
\end{equation}
In this representation of the theory the gravitational field equation
reduces to an original Self Creation cosmology (Barber, 1982) in which the
scalar field is minimally coupled to the metric, and which only interacts
with the material universe by determining the gravitational coefficient $G$.
This original representation of the theory has been the subject of some
discussion over the past twenty years, [Abdel-Rahman, (1992), Abdussattar \& Vishwakarma, 
(1997), Barber, (2002), (2003a), (2003b), Brans, (1987), Maharaj, (1988), Mohanty \& Mishra, (2002), Mohanty \& Mishra \& 
Biswal, (2002), Mohanty, Sahu \& Panigrahi, (2002), Mohanty, Sahu \& Panigrahi, (2003), 
Pimentel, (1985,) Pradhan \& Pandey (2002), Pradhan \& Vishwakarma (2002), Rahman \& Abdel,
 (1993), Ram \& Singh, (1998a), (1998b), Reddy, (1987a), (1987b), (1987c), (1987d),
Reddy, Avadhanulu \& Venkateswarlu, (1987), (1988), Reddy \& Venkateswarlu,(1988), (1989), 
Sahu \& Panigrahi, (2003), Sanyasiraju \& Rao, (1992), Shanthi \& Rao, (1991), Singh, 
Singh \& Srivastava, (1987), Singh \& Singh, (1984), Singh, (1986), (1989), 
Soleng, (1987a), (1987b), Venkateswarlu \& Reddy, (1988), (1989a), (1989b), (1989c), 
(1990), Wolf, (1988)].

\section{The Standard Formulae of SCC}

\subsection{The SCC Field Equations}

The SCC Action Principle gives rise to the following set of equations:

The scalar field equation 
\begin{equation}
\Box \phi =4\pi T_M^{\;}\text{ ,}  \label{eq143}
\end{equation}

The gravitational field equation

\begin{eqnarray}
R_{\mu \nu }-\frac 12g_{\mu \nu }R &=&\frac{8\pi }\phi T_{M\mu \nu }-\frac
3{2\phi ^2}\left( \nabla _\mu \phi \nabla _\nu \phi -\frac 12g_{\mu \nu
}g^{\alpha \beta }\nabla _\alpha \phi \nabla _\beta \phi \right)
\label{eq143a} \\
&&\ \ +\frac 1\phi \left( \nabla _\mu \nabla _\nu \phi -g_{\mu \nu }\Box
\phi \right) \text{ ,}  \nonumber
\end{eqnarray}

The creation equation, which replaces the conservation equation (Equation 
\ref{eq32a})

\begin{equation}
\nabla _\mu T_{M\;\nu }^{\;\mu }=\frac 1{8\pi }\frac 1\phi \nabla _\nu \phi
\Box \phi  \label{eq144}
\end{equation}
	These field equations are manifestly covariant, there is no preferred frame of reference
or absolute time. However in order to solve them one has to adopt a specific coordinate 
system; the CoM of the system in the spherically symmetric One Body Case, or that of the 
comoving fluid of the cosmological solution.  In those frames of reference there is a specific 
coordinate time as in the standard GR solutions.
    These JF(E) solutions of SCC, moreover, have the property not only of being in the Machian 
frame of reference but also of locally conserving mass-energy.

\subsection{The Spherically Symmetric Solution}

The Robertson parameters are 
\begin{equation}
\alpha _r=1\qquad \beta _r=1\qquad \gamma _r=\frac 13\text{ ,}  \label{eq145}
\end{equation}
and therefore the standard form of the Schwarzschild metric is 
\begin{eqnarray}
d\tau ^2 &=&\left( 1-\frac{3G_NM}r+..\right) dt^2-\left( 1+\frac{G_NM}%
r+..\right) dr^2  \label{eq146} \\
&&\ \ \ \ \ \ \ -r^2d\theta ^2-r^2\sin ^2\theta d\varphi ^2\text{ .} 
\nonumber
\end{eqnarray}
The formula for $\phi $ is 
\begin{equation}
\phi =G_N^{-1}\exp (-\Phi _N)  \label{eq108}
\end{equation}
and that for $m$ is, (Equation \ref{eq24}), 
\[
m_p\left( x_\mu \right) =m_0\exp (\Phi _N)\text{.}  
\]
Hence we note that in this case, in Equation \ref{eq34c}, $\alpha =-1$ and $%
\beta =-3$ , thus in the spherically symmetric solution to the field
equations Equation \ref{eq34d} becomes 
\begin{equation}
\Omega =\left( G\phi \right) ^{-1}\text{ .}  \label{eq24aa}
\end{equation}

\subsection{Local Consequences of the Theory}

There are two Gravitational constants, $G_N$, which applies to particles and 
measurable in Cavendish type experiments as the standard Newtonian constant
and $G_m$, which applies to photons and is that constant which determines
the curvature of space-time. These two constants relate together according
to 
\begin{equation}
G_N=\frac 23G_m\text{ .}  \label{eq149}
\end{equation}
Hence, if normal Newtonian gravitational acceleration is $g$, the
acceleration of a massive body caused by the curvature of space-time is $%
\frac 32g$ 'downward' compensated by an 'upward' acceleration caused by the
scalar field of $\frac 12g$.

Finally in the JF(E) the radial inward acceleration of a freely falling body
is given by the non-Newtonian expression 
\begin{equation}
\frac{d^2r}{dt^2}=-\left\{ 1-\frac{G_NM}r+...\right\} \frac{G_NM}{r^2}\text{
.}  \label{eq150}
\end{equation}
In the earlier paper it was seen that the effect of this non-Newtonian
perturbation was to compensate for the effect of the scalar field upon the
curvature of space-time.

The acceleration experienced by a freely falling particle is given by 
\begin{equation}
m_0\frac{d^2r}{dt^2}=-m(r)\frac{G_NM}{r^2}\text{ .}  \label{eq151}
\end{equation}
We see that $m_0$ can be thought of as 'inertial-mass', which measures
inertia and $m(r)$ as 'gravitational mass', which interacts with the
gravitational field with 
\[
\stackunder{r\rightarrow \infty }{Lim}\text{ }m(r)=m_{0\text{ }}\text{.} 
\]

As described in the original paper, (Barber, 2002), the conformal
equivalence between the JF(E) and the EF, which is canonical GR, results in
the predictions in the standard tests being identical with GR. In the JF(E)
it was seen in detail that the action of the non-conservation of the
energy-momentum tensor for matter resulted in an extra 'scalar-field' force
acting on particles which exactly compensated for the scalar field
perturbation of the curvature of the space-time manifold. Nevertheless two
definitive experiments were suggested which examine the interaction of the
photon and the vacuum energy fields with ordinary matter.

\section{The Cosmological Solution}

\subsection{Deriving the General Cosmological Equations}

Using the Cosmological Principle the usual assumptions of homogeneity and
isotropy can be made to obtain the cosmological solutions to the field
equations. 

The privileged CoM frame in which physical units may be defined for any
epoch is now the ''rest frame'' for the universe as a whole. Presumably it
should be identified physically with that frame in which the microwave
background radiation is globally isotropic.

According to SCC, a gravitational field, i.e. the curvature of space-time,
is to be described in the JF(E), whereas observations using atomic
apparatus, based on an atomic clock, are referred to the EF. The two frames
have to be transformed as appropriate.

There are four equations to consider. The first is the Gravitational Field
Equation \ref{eq143a}, which is exactly the same as the BD equation with $%
\omega =-\frac 32$ in the BD equation. The second is the Scalar Field
Equation \ref{eq2}. In GR the third equation is the conservation equation
which is replaced in SCC by the Creation Field Equation \ref{eq31}. The
fourth equation is some equation of state, such as the dust filled universe $%
p=0$, or the early radiation dominated universe in which $p=\frac 13\rho $.
The SCC field equations demand an exotic equation of state.

The two gravitational cosmological equations are 
\begin{equation}
\left( \frac{\stackrel{.}{R}}R\right) ^2+\frac k{R^2}=+\frac{8\pi \rho }{%
3\phi }-\frac{\stackrel{.}{\phi }\stackrel{.}{R}}{\phi R}-\frac 14\left( 
\frac{\stackrel{.}{\phi }}\phi \right) ^2\text{ ,}  \label{eq197}
\end{equation}
\begin{equation}
\frac{\stackrel{..}{R}}R+\left( \frac{\stackrel{.}{R}}R\right) ^2+\frac
k{R^2}=-\frac 16\left( \frac{\stackrel{..}{\phi }}\phi +3\frac{\stackrel{.}{%
\phi }\stackrel{.}{R}}{\phi R}\right) +\frac 14\left( \frac{\stackrel{.}{%
\phi }}\phi \right) ^2\text{ .}  \label{eq199}
\end{equation}

The scalar cosmological equation 
\begin{equation}
\stackrel{..}{\phi }+\,3\frac{\stackrel{.}{\phi }\stackrel{.}{R}}R=4\pi
\left( \rho -3p\right) \text{ .}  \label{eq202}
\end{equation}

The creation cosmological equation is 
\begin{equation}
\stackrel{.}{\rho }\,=-3\frac{\stackrel{.}{R}}R\left( \rho +p\right) +\frac
1{8\pi }\frac{\stackrel{.}{\phi }}\phi \left( \stackrel{..}{\phi }+\,3\frac{%
\stackrel{.}{\phi }\stackrel{.}{R}}R\right) \text{ .}  \label{eq204}
\end{equation}
(It is a moot point whether the scalar field $\phi $ is generated by the
distribution of mass and energy via Equation \ref{eq202}, or whether mass is
generated by the scalar field via Equation \ref{eq204}.)

Finally the equation of state remains 
\begin{equation}
p=\sigma \rho \text{ ,}  \label{eq205}
\end{equation}
where $\sigma =+\frac{1}{3}$ in a radiation dominated universe and $\sigma
=0 $ in a dust filled universe.

\subsection{The SCC Cosmological Solution and Consequences}

The five independent Equations, \ref{eq197}, \ref{eq199}, \ref{eq202}, \ref
{eq204} and \ref{eq205} and the sixth relationship, provided by the
conservation of a free photon's energy in the JF(E) together with Stephan's
Law, provide a solution for the six unknowns $R(t)$, $\phi (t)$, $\rho (t)$, 
$p(t)$, $k$ and $\sigma $. There are also the boundary conditions at $t=t_0$
(present epoch), $R_0$, $\phi _0$, $\rho _0$, and $p_0$.

The cosmological 'self-creation equation' is found to be 
\begin{equation}
\rho =\rho _0\left( \frac R{R_0}\right) ^{-3\left( 1+\sigma \right) }\left(
\frac \phi {\phi _0}\right) ^{\frac 12\left( 1-3\sigma \right) }\text{ ,}
\label{eq206}
\end{equation}
which is the equivalent GR expression with the addition of the last factor
representing cosmological 'self-creation'. However for a photon gas $\sigma
=+\frac 13$ so Equation \ref{eq206} reduces to its GR equivalent, consistent
with the Principle of Mutual Interaction that there is no interaction
between a photon and the scalar field, 
\begin{equation}
\rho _{em}=\rho _{em\,0}\left( \frac R{R_0}\right) ^{-4}\text{ .}
\label{eq207}
\end{equation}
Since $\rho _{em}\propto T_{em}^4$ where $T_{em}$ is the Black Body
temperature of the radiation, the GR relationship $T_{em}\propto R^{-1}$
still holds. Also as the wavelength $\lambda _{em}$ of maximum intensity of
the Black Body radiation is given by $\lambda _{em}\propto T_{em}^{-1}$, SCC
JF(E) retains the GR relationship 
\begin{equation}
\lambda _{em}\propto R\text{ .}  \label{eq208}
\end{equation}

However in the SCC JF(E) $\lambda _{em}$ is constant for a free photon, even
over curved space-time, and it is particle masses which vary. Therefore in
the JF(E) Equation \ref{eq208} becomes simply 
\begin{equation}
R=R_0\text{ .}  \label{eq209}
\end{equation}
In the Jordan energy frame the universe is static when measured by light,
that is a co-expanding ''light ruler'' is unable to detect the expanding
universe.

The cosmological gravitational and scalar field equations are solved to
yield 
\begin{equation}
\left( 5-3\sigma \right) \frac{\stackrel{..}{\phi }}\phi =3\left( 1-3\sigma
\right) \left( \frac{\stackrel{.}{\phi }}\phi \right) ^2  \label{eq212}
\end{equation}
which has the two possible solutions;

Case 1 when $\sigma \neq -\frac 13$%
\begin{equation}
\phi =\phi _0\left( \frac t{t_0}\right) ^{-2}\text{ ,}  \label{eq213}
\end{equation}
which corresponds to a universe empty of everything except the false vacuum.

The presence of any matter or electromagnetic energy in the universe forces
the solution to assume Case 2 with 
\begin{equation}
\sigma =-\frac 13\text{ .}  \label{eq219}
\end{equation}
In which case Equation \ref{eq212} has been shown (Barber, 2002) to have the
solution 
\begin{equation}
\phi =\phi _0\exp \left( H_0t\right) \text{ ,}  \label{eq214}
\end{equation}
where $H_0$ is Hubble's 'constant' in the present epoch, defined by $t=0$ ,
and $\phi _0=G_N^{-1}$. By definition $G_N$ is the value measured in
''Cavendish type'' experiments in the present epoch. Note the theory admits
a cosmological ground state solution, $g_{\mu \nu }\rightarrow \eta _{\mu
\nu }$ and $\nabla _\mu \phi =0$ only when $t\rightarrow -\infty $, that is
at the ''Big Bang'' itself. Equations \ref{eq202}, \ref{eq205}, \ref{eq209}, 
\ref{eq214} and \ref{eq219} yield 
\begin{equation}
\frac{8\pi \rho }{\phi _0}=H_0^2\exp \left( H_0t\right) \text{ .}
\label{eq221}
\end{equation}
This can be written in the form 
\begin{equation}
\rho =\rho _0\exp \left( H_0t\right) \qquad  \label{eq222}
\end{equation}
\begin{equation}
\text{where\ }\rho _0=\frac{H_0^2}{8\pi G_N}\text{ ,}  \label{eq238}
\end{equation}
if now, as usual, the critical density is defined $\rho _c=\frac{3H_0^2}{%
8\pi G_N}$, then $\rho _0=\frac 13\rho _c$. Hence the cosmological density
parameter $\Omega _c$ 
\begin{equation}
\Omega _c=\frac 13\text{ . }  \label{eq239}
\end{equation}
Therefore, in this theory there is no need for 'Dark Energy'. The
cosmological density parameter $\Omega _c$ comprises of baryonic (plus any
cold dark matter) and radiation (plus any hot dark matter) components
together with that of false vacuum energy. As the total pressure is
determined by the constraints of the cosmological equations Equation \ref
{eq219}, together with Equation \ref{eq222}, the total cosmological pressure
is given by 
\begin{equation}
p=-\frac 13\rho _0\exp \left( H_0t\right) \text{ .}  \label{eq241}
\end{equation}
To explain this it is suggested that a component of the cosmological
pressure and density is made up of false vacuum. That is there is a
''remnant'' vacuum energy made up of contributions of zero-point energy from
every mode of every quantum field which would have a natural energy
''cut-off'' $E_{\max }$ which in the cosmological case is determined, and
limited, by the solution to the cosmological equations. Let there be three  
species, baryons, electromagnetic radiation and false vacuum: 
\begin{equation}
p_b+p_{em}+p_f=-\frac 13\left( \rho _b+\rho _{em}+\rho _f\right) \text{ ,}
\label{eq249}
\end{equation}
and as in the present epoch $p_b\approx 0$, $\rho _{em}\approx 0$, $%
p_{em}\approx 0$, and $p_f=-\rho _f$ we are left with 
\begin{equation}
\rho _b=2\rho _f\text{ .}  \label{eq250}
\end{equation}
Therefore the density parameter for cold matter (visible and dark) is 
\begin{equation}
\Omega _b=\frac 29\approx 0.22\ \text{.}  \label{eq251}
\end{equation}
the difference between $\Omega _c$ and $\Omega _b$ would be interpreted as
the hot dark matter component of ''missing mass'' or, as this component has
negative pressure and evolves with time, it might be presently identified
with ''quintessence'' [Cruz, N. et al., (1998)], [Huey, G. et al., (1999)],
[Zlatev, I. et al., (1999)]. As this component is determined by the field
equations the 'lambda problem' may have been resolved.

Assuming baryon conservation in a static universe, the inertial mass of a
fundamental particle must be given by 
\begin{equation}
m_i=m_0\exp \left( H_0t\right) \text{ .}  \label{eq223}
\end{equation}

A GR expanding universe with constant atomic masses of invariant size is
replaced in SCC by a static universe with increasing atomic masses of
decreasing size, that is ''shrinking rulers''.

\subsection{The Transformation Into the Einstein Frame (EF)}

Measurements of curvature or the wavelength/energy of a photon are made in
the JF(E), however the physics of atomic structures is naturally described
in the EF. It is now necessary to transform the units used in the JF(E) into
the system used in physical measurement using atomic apparatus, that is the
EF. The two frames are conformally related by Equation \ref{eq9}, using $%
\Omega $ again as the parameter of conformal transformation, 
\[
g_{\mu \nu }\rightarrow \widetilde{g}_{\mu \nu }=\Omega ^2g_{\mu \nu }\text{
,} 
\]
where the interval is invariant under the transformation

\begin{equation}
d\tau ^2=-g_{\mu \nu }dx^\mu dx^\nu =-\widetilde{g}_{\mu \nu }d\widetilde{x}%
^\mu d\widetilde{x}^\nu \text{ . }  \label{eq224}
\end{equation}

Now mass transforms according to Equation \ref{eq33}

\[
m\left( x^\mu \right) =\Omega \widetilde{m}\text{ ,} 
\]
therefore Equation \ref{eq223} requires in the cosmological solution to the
field equations 
\begin{equation}
\Omega =\exp \left( H_0t\right) \text{ .}  \label{eq225}
\end{equation}
The comparison of Equation \ref{eq222} with Equation \ref{eq214} reveals
that $\rho \propto \phi $ . Note that although in the One Body Problem $%
m\left( r\right) \varpropto \phi \left( r\right) ^{-1}$, cosmologically $%
m\left( t\right) \varpropto \phi \left( t\right) $ . Hence from Equation \ref
{eq34c}, in the cosmological solution to the field equations, $\alpha =+1$
and $\beta =-1$ and Equation \ref{eq34d} becomes 
\begin{equation}
\Omega =G\phi \text{ .}  \label{eq223a}
\end{equation}
From which length and time transform, by integrating along spacelike and
timelike paths respectively 
\begin{equation}
\widetilde{L}=L_0\exp \left( H_0t\right) \text{ }  \label{eq226}
\end{equation}
\begin{equation}
\text{and\ }\triangle \widetilde{t}=\triangle t\exp \left( H_0t\right) \text{
.}  \label{eq227}
\end{equation}
These transformations are consistent with using the Bohr/Schr\"odinger/Dirac
models of an atom to measure length and time under mass transformation.

The two time scales relate to each other as follows 
\begin{equation}
\widetilde{t}=\frac 1{H_0}\exp \left( H_0t\right) \text{\qquad and \qquad }%
t=\frac 1{H_0}\ln \left( H_0\widetilde{t}\right) \text{ ,}  \label{eq227a}
\end{equation}
where $\widetilde{t}$ is time measured from the ''Big Bang'' in the EF, and $%
t$ is time measured from the present day in the JF(E).

Applying this transformation to the universe's scale factor in two steps,
the first step yields 
\begin{equation}
\widetilde{R}=R_0\exp \left( H_0t\right) \text{ .}  \label{eq228}
\end{equation}
This expression uses mixed frames, that is length is in the EF and time is
in the JF(E).If we now substitute for $t$ in Equation \ref{eq228} we obtain
the scale factor of the universe in the EF. 
\begin{equation}
\widetilde{R}=R_0\frac{\widetilde{t}}{\widetilde{t}_0}\text{ .}
\label{eq233}
\end{equation}

Thus when measured by physical rulers and clocks the universe is seen to
expand linearly from a ''Big Bang''. The deceleration parameter 
\begin{equation}
q=-\left( \frac{\stackrel{..}{\widetilde{R}}}{H^2\widetilde{R}}\right) =0%
\text{ .}  \label{eq234}
\end{equation}
Therefore the horizon, smoothness and density problems of classical GR
cosmology, which all arise from a positive, non zero $q$, do not feature in
SCC. Hence it is unnecessary to invoke Inflation in this theory and indeed,
with Equation \ref{eq228}, SCC might be considered to be a form of
''Continuous Inflation''.

The curvature constant $k$ is given by Equations \ref{eq199} and \ref{eq197} 
\begin{equation}
\frac k{R_0^2}=+\frac 1{12}H_0^2\text{ ,}  \label{eq242}
\end{equation}
so $k$ is positive definite, 
\begin{equation}
k=+1\text{ ,}  \label{eq243}
\end{equation}
that is the universe is finite and unbounded. From Equation \ref{eq242} $R_0$
can be derived in terms of the Hubble time 
\begin{equation}
R_0=\sqrt{12}H_0^{-1}\text{ .}  \label{eq244}
\end{equation}

This may be seen by comparing Equation \ref{eq197} 
\begin{equation}
\left( \frac{\stackrel{.}{R}}R\right) ^2=+\frac{8\pi \rho }{3\phi }-\frac
k{R^2}-\frac{\stackrel{.}{\phi }\stackrel{.}{R}}{\phi R}-\frac 14\left( 
\frac{\stackrel{.}{\phi }}\phi \right) ^2\text{ }  \label{eq251a}
\end{equation}
with its general GR equivalent 
\begin{equation}
\left( \frac{\stackrel{.}{R}}R\right) ^2=+\frac{8\pi G_N\rho }3-\frac
k{R^2}+\frac \Lambda 3\text{ .}  \label{eq251b}
\end{equation}
Thus a GR interpretation of a SCC universe would yield a cosmological
constant 
\begin{equation}
\Lambda =-3\left[ \frac{\stackrel{.}{\phi }\stackrel{.}{R}}{\phi R}+\frac
14\left( \frac{\stackrel{.}{\phi }}\phi \right) ^2\text{ }\right] \text{ .}
\label{eq251c}
\end{equation}
Now using the standard matter, curvature and cosmological constant density
parameters, 
\begin{equation}
\Omega _m=\frac{8\pi G_N\rho _0}{3H_0^2}\text{ , }\Omega _k=-\frac
k{R_0^2H_0^2}\text{ and }\Omega _\Lambda =\frac \Lambda {3H_0^2}\text{ , }
\label{eq251d}
\end{equation}
and the SCC solutions given by Equations \ref{eq238}, \ref{eq243}, \ref
{eq209}, \ref{eq244} and \ref{eq214} we obtain 
\begin{equation}
\Omega _m=\frac 13\text{ , }\Omega _k=-\frac 1{12}\text{ and }\Omega
_\Lambda =-\frac 14\text{ , i.e. }\Omega _m+\text{ }\Omega _k+\Omega
_\Lambda =0\text{ .}  \label{eq251e}
\end{equation}
This explains the static universe in the JF(E) as in a GR analysis the
energy contributions of matter, curvature and cosmological constant cancel
out, in a similar way to the original static Einstein model. Furthermore it should be noted 
that the above value $\Omega_\Lambda = -\frac 14$ cannot be observed directly. At present, 
using a model based on GR, the total energy parameter is deduced from the observed spatial 
flatness to be unity. Therefore if the universe is as predicted by SCC so that 
$\Omega_m = +\frac 13$ then it would be thought that the density due to the cosmological constant 
is $\Omega_\Lambda = +\frac 23$, as indeed is the case.  

\subsection{Mixed-Frame Measurements}

The frame in which a cosmological observation is made has to be carefully
considered. Atomic measurements lend themselves to the EF, while remote
observations receiving photons lend themselves to the JF(E). Some
cosmological observations may be comparing two quantities each measured in
either frame. For example, if the ''expansion'' of the universe is expressed
comparing EF length and JF(E) time we obtain Equation \ref{eq228} 
\[
\widetilde{R}=R_0\exp \left( H_0t\right) \text{ .} 
\]

Observations of distant supernovae compare the atomic theory of stellar
evolution and supernova luminosity - an assessment made in the EF, with red
shift, a geometric measurement made in the JF(E). In the SCC it is expected
that such observations would detect a universe expanding exponentially
according to Equation \ref{eq228}.

The mixed-frame expression for $\phi $ is given by Equations \ref{eq227a},
and \ref{eq214} 
\begin{equation}
\phi =\frac 1{G_N}\text{ }\frac{\widetilde{t}}{t_0}\text{ }\backsim 
\widetilde{t}\text{ .}  \label{eq245}
\end{equation}
This might explain the Large Numbers Hypothesis (LNH) relationship $G\approx
T^{-1}$ where $G$ and $T$ are the normal LNH dimensionless values of the
gravitational ''constant'' and the age of the universe respectively.

By definition the mass of a fundamental particle in the EF, $\widetilde{m}$,
is constant, although when measured by comparison with the JF(E) energy of a
free photon, the mixed frame mass bears the same linear relationship 
\begin{equation}
m\left( \widetilde{t}\right) =m_0\text{ }\frac{\widetilde{t}}{t_0}=m_0\frac{%
\widetilde{R}}{R_0}\text{ ,}  \label{eq246}
\end{equation}
which is normally interpreted in the EF as the free photon suffering a red
shift 
\begin{equation}
1+z=\frac{R_0}{\widetilde{R}}\text{ .}  \label{eq247}
\end{equation}
Cosmological redshift in a static universe is interpreted as a measurement
of the cosmological increase of the atomic masses of the measuring apparatus rather than 
by a 'doppler shift'. Also as a check, Equations \ref{eq245} and \ref{eq246} give the 
mixed frame variation of $\widetilde{%
\phi }$ as 
\begin{equation}
\phi \left( \widetilde{t}\right) =\frac 1{G_N}\text{ }\frac m{m_0}\text{ ,}
\label{eq248}
\end{equation}
\[
\text{so\qquad }G\left( \widetilde{t}\right) m\left( \widetilde{t}\right)
=G_Nm_0=\text{ a constant.} 
\]

This confirms that if atomic masses are the standard of mass and are thereby
defined to be invariant in the EF, $\widetilde{\phi }$ and hence $G$ are
necessarily invariant also. Therefore whether $G$ is observed to vary or not
will depend on which frame is used to interpret the results. When measured
in the EF, the gravitational field of a massive body remains invariant over
cosmological time.

\section{Conclusions}

The theory explains the present quandary about the observed density of the
universe and accelerating cosmological expansion. [Garnavich, P.M. et al.,
(1998)], [Filippenko, A.V. et al., (1998)], [Riess, A.G. et al., (1998)]
There is no need to invoke a cosmological constant, dark energy and/or
quintessence. There is, nevertheless, a requirement for dark matter of
around $\Omega _{dm}=0.2$ to explain galactic clustering. In the JF(E),
appropriate for observations of photons, the universe is similar to the
original Einstein static cylinder model. Hence it is spatially flat in
agreement with observations of CMB anisotropies even though the total
density parameter $\Omega _m=\frac 13$.

As $GM$ is invariant, and because of the conformal equivalence between SCC
and GR, gravitational orbits in the JF(E) are the same as in GR. However
there is a secular variation in measuring space and time. The EF unit of time 
is that measured by atomic clocks, whereas the JF(E) unit of time is that determined 
by the frequency of a 'standard' photon and gravitational orbits, i.e. ephemeris time. 
In SCC the anomalous acceleration of the Pioneer spacecraft is explained as a clock 
shift between ephemeris time (JF(E) clock) and atomic clock time (EF clock). 
(Anderson et al., 2002a) (also see Ostermann 2002).

An analysis of the residuals of planetary longitudes rendered a value for a
variation in $G$ 
\begin{equation}
\frac{\stackrel{.}{G\left( t\right) }}{G\left( t\right) }\approx +\left(
4\pm 0.8\right) .10^{-11}yr^{-1}  \label{eq248a}
\end{equation}
with a caveat that the sign might be reversed. [Krasinsky et al., (1985)]
On the other hand, they also reported the contradictory null result
[Hellings et al., (1983)] determined from accurate observations of the
Viking Landers and the Mariner 9 spacecraft. This null result has been
confirmed repeatedly since, recently by Anderson et al. and others (Anderson
et al., 2002b), (Williams, 2001). The discrepancy between these two results
might be explained by SCC, if a detailed analysis of these results
accordingly to SCC is carried out in the future. The discrepancy may depend
on whether $\frac{\stackrel{.}{G}}G$ is being measured or $\frac{\stackrel{.%
}{\left( GM\right) }}{GM}$. As stated above $GM$ is invariant in the theory.
The null Viking result may be explained by the clock drift between ephemeris time 
and universal (atomic clock) time being compensated by a secular evolution of the
spherically symmetric solution to the field equations.

Another observable effect arises in the JF(E) as a result of the variation
in $m(t)$. If angular momentum is conserved then $mr^2\omega $ is constant,
but as atomic masses increase secularly, their radii will shrink. 
\[
\text{If }m(t)=m_0\exp (H_0t)\text{ (Equation \ref{eq223}),} 
\]
\[
\text{ then }r(t)=r_0\exp (-H_0t)\text{ (The inverse of Equation \ref{eq228})%
} 
\]
\[
\text{and if }\frac d{dt}\left( mr^2\omega \right) =0\text{ ,} 
\]
\begin{equation}
\text{then }\frac{\stackrel{.}{\omega }}\omega =-\left[ \frac{\stackrel{.}{m}%
}m+2\frac{\stackrel{.}{r}}r\right] =+H_0\text{ ,}  \label{eq248b}
\end{equation}
and solid bodies such as the Earth should spin up when measured by JF(E),
(ephemeris) time. It has indeed been reported that this spin up is observed.
As mentioned above, the review by Leslie Morrison and Richard Stephenson
[(Morrison and Stephenson 1998),(Stephenson 2003)], studying the analysis of the length of the
day from ancient eclipse records reported that in addition to the tidal
contribution there is a long-term component acting to decrease the length of
the day which equals 
\[
\vartriangle \text{T/day/cy}=-6\ \text{x }10^{-4}\text{ sec/day/cy.} 
\]
This value, equivalent to $H=67$ $km.sec^{\text{-1}}/Mpc$, is remarkably close to the 
best estimates of $H_0$. However at least part of this spin up is probably caused by a 
decrease of the Earth's moment of inertia.

How significant are these anomalies? Is their proximity to the value of $H_0$
merely coincidence, or is there new physics here? If it is new physics that
is being observed then it is suggested here that SCC would be a candidate
worth consideration. Either way the situation may be clarified by performing
the definitive experiments described in principle in the earlier paper,
(Barber, 2002) if it is not resolved earlier by the Gravity Probe B experiment.   

\section{Acknowledgments}

I am thankful for discussion with Professor Bernard Carr, Queen Mary and
Westfield College, University of London. Any errors and misconceptions
remaining are of course entirely my own.

\end{document}